\documentclass[aps,10pt,reprint,groupedaddress,superscriptaddress]{revtex4-2}

\usepackage[usenames,dvipsnames]{xcolor}
\usepackage{amssymb}
\usepackage{graphicx}
\usepackage{amsmath}
\usepackage{enumitem}
\usepackage{txfonts}
\usepackage[bookmarks=true,colorlinks,citecolor=blue,urlcolor=blue]{hyperref}
\usepackage{braket}
\usepackage{babel}
\usepackage{blindtext}
\usepackage[normalem]{ulem}

\begin{document}

\title{Isometric tensor network representations of two-dimensional thermal states}
\author{Wilhelm Kadow}
\affiliation{Department of Physics, Technical University of Munich, 85748 Garching, Germany}
\affiliation{Munich Center for Quantum Science and Technology (MCQST), Schellingstr. 4, D-80799 M{\"u}nchen, Germany}
\author{Frank Pollmann}
\affiliation{Department of Physics, Technical University of Munich, 85748 Garching, Germany}
\affiliation{Munich Center for Quantum Science and Technology (MCQST), Schellingstr. 4, D-80799 M{\"u}nchen, Germany}
\author{Michael Knap}
\affiliation{Department of Physics, Technical University of Munich, 85748 Garching, Germany}
\affiliation{Munich Center for Quantum Science and Technology (MCQST), Schellingstr. 4, D-80799 M{\"u}nchen, Germany}
\date{\today}

\begin{abstract}
Tensor networks provide a useful tool to describe low-dimensional complex many-body systems. Finding efficient algorithms to use these methods for finite-temperature simulations in two dimensions is a continuing challenge. Here, we use the class of recently introduced isometric tensor network states, which can also be directly realized with unitary gates on a quantum computer. We utilize a purification ansatz to efficiently represent thermal states of the transverse field Ising model. By performing an imaginary-time evolution starting from infinite temperature, we find that this approach offers a different way with low computational complexity to represent states at finite temperatures. 
\end{abstract}

\maketitle
    
\section{Introduction}

Quantum many-body systems can form fascinating phases of matter with exotic emergent properties~\cite{Wen2017}. Discovering these phases and characterizing their stability to thermal fluctuations is a pertinent challenge, especially since any experiment will naturally be conducted at finite temperature. 
To obtain finite-temperature properties of generic interacting quantum many-body systems numerical techniques are typically required. Quantum Monte Carlo (QMC) sampling of finite temperature states is extremely efficient for many systems. However, fermionic and frustrated models cannot be sampled due to the infamous sign problem~\cite{Sandvik2010}. For these systems exact diagonalization can be applied, which is restricted to very small systems, because of the exponential growth of the Hilbert space. An alternative approach is provided by tensor network states (TNS)~\cite{Fannes1992, Nishino1996, Nishino1998, Verstraete2004a, Banuls2008, Cirac2021}. These variational states can capture low-entanglement regions of the full Hilbert space. Proven rigorously for gapped Hamiltonians in one dimension~\cite{Hastings2007} and for general thermal states~\cite{Wolf2008}, the area law claims, that the entanglement for the ground state of such a Hamiltonian is only proportional to the area between two subsystems.

For one-dimensional systems TNS have been very successful~\cite{White1992, Vidal2004, Schollwoeck2011, Tenpy}. There the area law translates to a constant entanglement between two subsystems independent of their size. Several algorithms for matrix product states (MPS) can be used to efficiently obtain not only ground states but also thermal states by an imaginary-time evolution of purified~\cite{Verstraete2004, Feiguin2005, Barthel2009} or minimally entangled typical thermal states~\cite{White2009, Binder2015}. Key insight comes from the gauge degree of freedom of MPS, which allows one to write them in a canonical form. This imprints an isometry conditions on the MPS. That in turn enables for an efficient computation of local observables.
In two dimensions, TNS are difficult to handle numerically as the contraction of these networks is in general exponentially hard~\cite{Verstraete2006, Schuch2007}, so one usually has to resort to approximate contraction schemes~\cite{Nishino1996, Jordan2008, Jiang2008, Lubasch2014, Lubasch2014a}. Recently a subclass of two-dimensional (2D) TNS has been introduced which obeys a similar isometry condition as one-dimensional (1D) MPS and are hence called isometric tensor network states (isoTNS)~\cite{Zaletel2020, Haghshenas2019, Hyatt2019, Soejima2020, Tepaske2021}. In~\cite{Zaletel2020} it was shown how to generalize the well known time-evolving block decimation (TEBD) algorithm from MPS to isoTNS and~\cite{Lin2021} adapts the density matrix renormalization group (DMRG) to approximate ground states with isoTNS. Very recent works also extend the isometric concept to fermionc~\cite{Dai2022} and semi-infinite tensor networks~\cite{Wu2022}.
Another convenient property of these states is that they can be directly implemented via unitary gates and hence can be represented on quantum computers~\cite{Wei2022, Slattery2021}. Currently, it is actively explored how useful the restricted variational isoTNS manifold is as a new set of variational states compared to full TNS. 
Another prevailing question is to extend TNS methods to finite temperatures, which has recently successfully been done for several models in 2D~\cite{Czarnik2014, Czarnik2015, Czarnik2016, Czarnik2019, Kshetrimayum2019, Chen2018}.

\begin{figure}[t]
\centering
\includegraphics[trim={0cm 0cm 0cm 0cm},clip,width=0.95\linewidth]{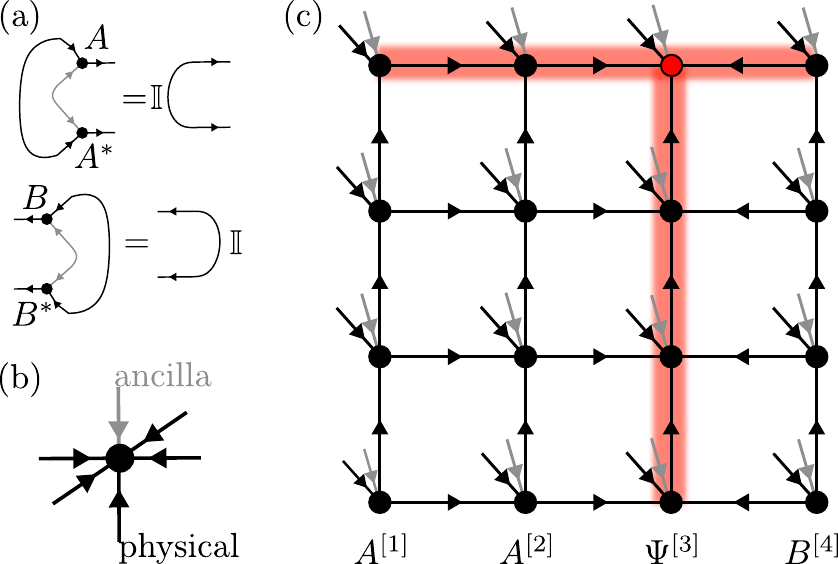}
\caption{\textbf{Purified isometric tensor networks.} (a) The isometry condition for left- and right-normalized tensors [see eqs. \eqref{eq::canonicalA} and \eqref{eq::canonicalB}] gives identities upon contraction of all ingoing legs of the state and its conjugate. (b) The local Hilbert space is doubled, illustrated by an ancilla leg (gray), to obtain a purified thermal state. (c) Two-dimensional purification of isoTNS with an orthogonality center (red tensor) and orthogonality hypersurface on the corresponding row and column (red shaded region). We label the left- and right-normalized columns $A^{[i]}$ and $B^{[i]}$ and the orthogonality column $\Psi^{[i]}$.}
\label{fig::1}
\end{figure}

In this work, we show that isoTNS can be used to efficiently compute finite-temperature properties for 2D systems. In particular, we focus on the transverse field Ising model, which undergoes a thermal ($T > 0$) as well as a quantum phase transition ($T=0$). Since this model is sign-problem free, we benchmark our results with QMC simulations. Yet, we emphasize that TNS methods are not affected by any sign problem and can therefore be used to simulate frustrated or fermionic systems as well.
\begin{figure*}[t]
\centering
\includegraphics[width=0.9\linewidth]{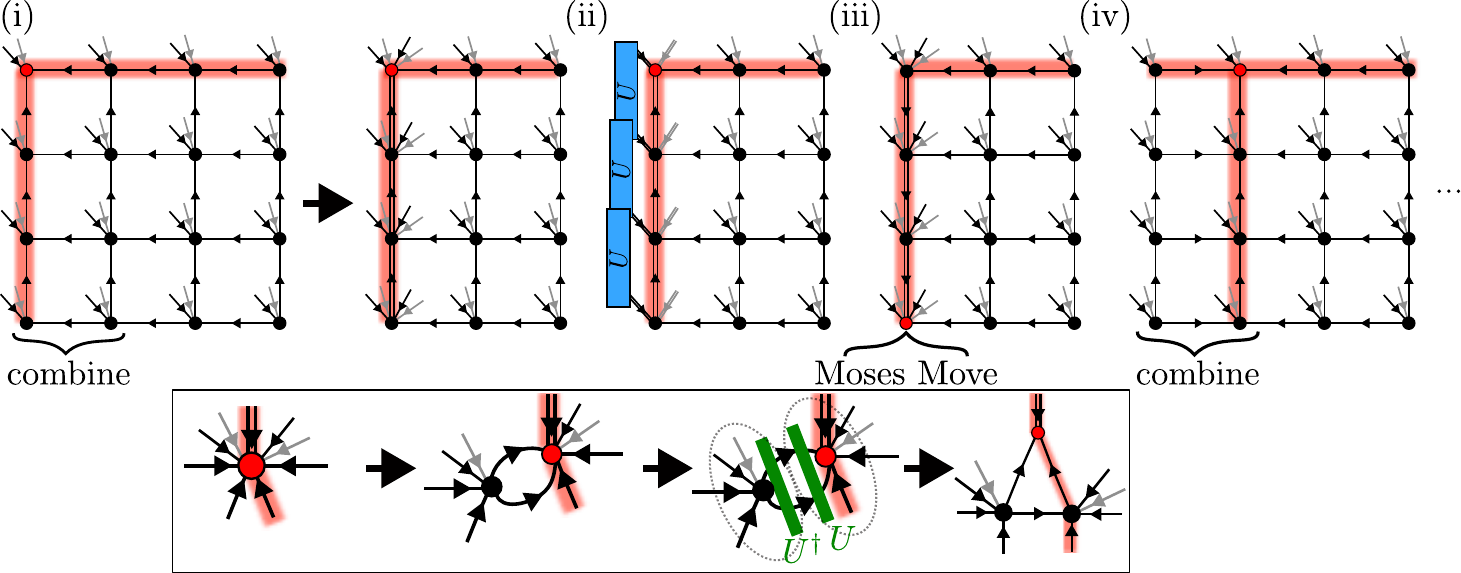}
\caption{\textbf{TEBD$^2$ algorithm for purification.} We use a slight variation of the TEBD$^2$ algorithm, presented in earlier studies~\cite{Zaletel2020, Dai2022, Wu2022}, where we combine two columns before applying the Trotter unitaries (blue). Steps (i)-(iv) illustrate the updates of the isoTNS within one sweep and are described in more detail in the main text. The inset shows the Moses Move, used to shift the orthogonality center from one column to the next one. The unitary gauge degree of freedom, used to disentangle the state, is depicted in green.}
\label{fig::2}
\end{figure*}

This work is structured as follows. In Sec.~\ref{sec::methods} we review the basic properties of isoTNS and an adapted version of the TEBD$^2$ algorithm used for the imaginary-time evolution to simulate thermal states via a purification ansatz. To test the capabilities of the algorithm we apply the algorithm to the 2D transverse-field Ising model and compare our results with QMC data in Sec.~\ref{sec::results}. We conclude in Sec.~\ref{sec::conclusion} and discuss the complexity of variational isoTNS represented on quantum computers.

\section{Imaginary-time evolution for purified \lowercase{iso}TNS}\label{sec::methods}
We now summarize the defining properties of isoTNS and describe the algorithm used to obtain thermal states in the purification picture.

The tensors of isoTNS are restricted to isometries, i.e., for $W\in \mathbb{C}^{n\times p}$, $n\geq p$, we have $W^\dagger W = \mathbb{I}_p$ and $WW^\dagger = \mathcal{P}_n$, where $\mathbb{I}_p$ is the identity on the smaller subspace and $\mathcal{P}_n$ is a projection from the larger subspace into the smaller one.
To include these properties in the graphical notation for the tensor networks we will write $W$ as a map $W:$ $\mathbb{C}^n(=\mathbb{C}^{n_1}\otimes \mathbb{C}^{n_2} \dots) \rightarrow \mathbb{C}^p(=\mathbb{C}^{p_1}\otimes \mathbb{C}^{p_2} \dots) $ and denote the domain (image) with ingoing (outgoing) arrows, respectively (see Fig.~\ref{fig::1}). Using the isometric form we can exactly contract the whole tensor network with its conjugate to a single site, which has only ingoing arrows. This site is referred to as orthogonality center.

In one dimension the celebrated canonical form of the MPS automatically implies these isometry constraints, such that we can write a state $\ket{\psi}$ in a general isometric form: 
\begin{equation}\label{eq::MPS}
    \ket{\psi} = \sum_{\substack{j_1, \cdots, j_L \\ \alpha_1, \cdots, \alpha_{L+1} }} A^{[1]j_1}_{\alpha_1 \alpha_2} A^{[2]j_2}_{\alpha_2 \alpha_3} \cdots \Psi^{[R_c]j_{R_c}}_{\alpha_{R_c} \alpha_{R_c +1}} \cdots B^{[L]j_L}_{\alpha_L \alpha_{L+1}} \ket{j_1 \cdots j_L}
\end{equation}
The physical sites are denoted by $j_i \in\lbrace 1,\cdots,d\rbrace$ for a \mbox{$d$-dimensional} local Hilbert space, while the virtual bond indices are labeled $\alpha_i~\in~\lbrace 1,\cdots,\chi\rbrace$ with the maximal bond dimension $\chi$. The isometric conditions for the left- and right-normalized $A$ and $B$ tensors are
\begin{align}
\sum_{j_i, \alpha_i} A^{[i]j_i}_{\alpha_i, \alpha_{i+1}} \overline{A}^{[i]j_i}_{\alpha_i, \overline{\alpha}_{i+1}} &= \delta_{\alpha_{i+1}, \overline{\alpha}_{i+1}} \label{eq::canonicalA} \\
\sum_{j_i, \alpha_{i+1}} B^{[i]j_i}_{\alpha_i, \alpha_{i+1}} \overline{B}^{[i]j_i}_{\overline{\alpha}_i, \alpha_{i+1}} &= \delta_{\alpha_{i}, \overline{\alpha}_{i}}, \label{eq::canonicalB}
\end{align}
which are also depicted in Fig.~\ref{fig::1}~(a). In general we can write an MPS in a form as in eq.~\eqref{eq::MPS} with the orthogonality center $\Psi^{[R_c]j_{R_c}}$ at site $R_c$. We can move the orthogonality center to a neighboring site and remain in isometric form by a QR decomposition or similar decompositions. Because of the isometric conditions for all tensors left and right of $R_c$, any local observable can be computed by contracting just the orthogonality center with the corresponding operator $O_{R_c}$:
\begin{equation}
    \bra{\psi} O_{R_c} \ket{\psi} = \sum_{\substack{j_{R_c}, \overline{j}_{R_c} \\ \alpha_{R_c}, \alpha_{R_c + 1}, } } O_{R_c}^{j_{R_c}, \overline{j}_{R_c}} \Psi^{[R_c]j_{R_c}}_{\alpha_{R_c}, \overline{\alpha}_{R_c+1}} \overline{\Psi}^{[R_c]\overline{j}_{R_c}}_{\alpha_{R_c}, \overline{\alpha}_{R_c+1}}.
\end{equation}
It is precisely the isometric form, that allows for very efficient algorithms. Updates of the local orthogonality center $\Psi$ are optimal, since all surrounding tensors in the environment can be contracted analytically.
Note that in one dimension we can always use the gauge degree of freedom to bring any MPS with fixed bond dimensions into isometric form.

In two dimensions the situation is different. In general the exact contraction of TNS is exponentially hard~\cite{Verstraete2006, Schuch2007}, which makes it also very difficult to find the optimal update of local tensors in time evolution or ground state search algorithms. The isometric conditions then help to contract the full tensor network in order to perform optimal tensor updates for two-dimensional versions of the DMRG and TEBD algorithm, named accordingly DMRG$^2$ and TEBD$^2$~\cite{Zaletel2020, Lin2021}.
To resemble the MPS notation, one keeps the form of the wave function as in Eq.~\eqref{eq::MPS}, but now $\Psi^{[i]}$ denotes an orthogonality column and the left- and right-normalized columns we label $A^{[i]}$ and $B^{[i]}$, respectively [see Fig.~\ref{fig::1}~(c)].
Below we will use the TEBD$^2$ algorithm for imaginary-time evolution of an infinite temperature state. Compared to the algorithm originally introduced in~\cite{Zaletel2020}, we use a slight variation, which was also applied in other recent works~\cite{Dai2022, Wu2022}, and is schematically shown in Fig.~\ref{fig::2}. Here, we summarize the most important steps and refer for a more detailed description to Refs.~\cite{Zaletel2020, Lin2021}:
\begin{enumerate}[label=(\roman*)]
    \item We start with an isoTNS with orthogonality center in the top left corner. This defines an orthogonality row and column where all arrows are ingoing. If we focus only on the column, we can think of it as an MPS, where the additional legs are included into the physical leg. Moving the orthogonality center from the top to the bottom corresponds in the MPS language to changing between a left- and right-canonical form. Before applying the Trotter gates for the imaginary-time evolution, we combine the orthogonality column with the next one. This leads to a new orthogonality column $\Psi B = \Psi''$, with increased virtual and physical bond dimensions.

    \item Using this double column representation, depicted by two parallel lines in the tensor network graph, we can apply the nearest-neighbor Trotter gates in horizontal and vertical direction at the same time. By updating the double-column tensors in a staircase fashion, the orthogonality center is brought from the top left to the bottom left corner. This scheme guarantees a second-order Trotterization error after sweeping back. Compared to the usual TEBD update scheme, where even and odd bonds are updated alternately, the staircase update brings the orthogonality center  automatically to the proper site for proceeding with the next step of the algorithm.

    \item To split the double column $\Psi ''$ again into two legs and to simultaneously bring the orthogonality center to the top of the next column, a ``Moses Move'' (MM) was proposed in~\cite{Zaletel2020}. This Moses Move corresponds to finding a left-normalized column $A$ and a new orthogonality center $\Psi$, such that the overlap with the initial $\Psi''$ is maximized,
    \begin{equation}\label{eq::costfunction}
        \underset{A, \Psi}{\mathrm{argmin}}\, ||\Psi'' - A\Psi||.
    \end{equation}
    To preserve the isometric form this can be done by either sequentially splitting up $\Psi''$ from the bottom to top or by variationally optimizing eq.~\eqref{eq::costfunction}. In practice it was found that the sequential splitting followed by some variational optimization steps gives the best results. During the sequential MM we can optimize the splitting to minimize the entanglement by utilizing a unitary disentangler [see inset in Fig.~\ref{fig::2}]. This was systematically investigated in~\cite{Lin2021} to optimize different entanglement measures. Here we will always use a non-linear disentangler, which optimizes the R\'enyi-1/2 entropy over the corresponding Riemannian manifold, as implemented in the \texttt{Pymanopt} package~\cite{Pymanopt}. Moreover, it was found in~\cite{Zaletel2020} that keeping an enlarged bond dimension $\eta$ on the vertical bonds of the new orthogonality column, decreases the error of the Moses Move substantially. Throughout this work we use ${\eta=2\,\chi}$.

    \item After splitting with the MM, we now combine the new orthogonality column with the next right-normalized column again to get a new double-column orthogonality center. Then we proceed by alternating TEBD, Moses and combining moves, until the orthogonality center has reached the top-right of the isoTNS. To reach the initial isometry configuration again, we perform all steps the other way around and therefore sweep the orthogonality column from the right back to the left. To do so we again use the double-column scheme to update horizontal and vertical bonds at the same time.
    
\end{enumerate}
\begin{figure*}[t!]
\centering
\includegraphics[trim={0cm 0cm 0cm 0cm},clip,width=\linewidth]{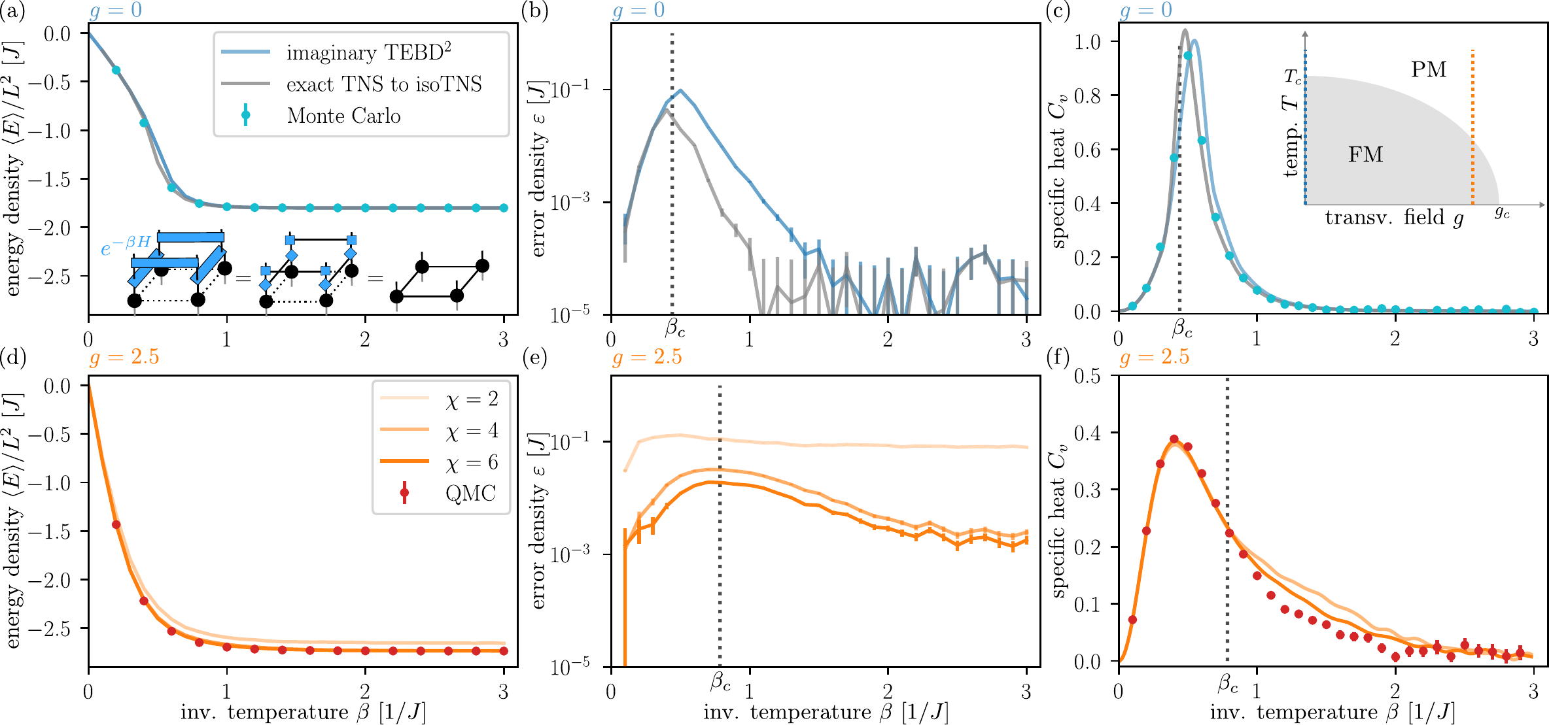}
\caption{\textbf{Energy densities of isoTNS.} The energy densities are obtained for the transverse field Ising model. A sketch of the phase diagram from a ferromagnet (FM) to a paramagnet (PM) shown as inset of (c). For the classical ($g=0$, blue lines) and quantum ($g=2.5$, orange lines) case we compare to (quantum) Monte Carlo results (QMC). For $g=0$ energies obtained from TEBD$^2$ with $\chi=4$ are shown in comparison with the exact TNS representation [inset in (a)] following an ``isometrization'' with Moses Move sweeps to an isoTNS with $\chi=4$. For $g=2.5$ we plot results for different maximal bond dimensions $\chi$. (a),~(d) Show the thermal energy density and (b),~(e) depict the error density $\varepsilon=|\langle E \rangle-E_{QMC}|/L^2$ compared to QMC results. Shown error bars arise from QMC sampling. By cubic spline interpolation of the thermal energy and computing the derivative numerically, we obtain the specific heat (c),~(f). All data are for $L=10$.}
\label{fig::3}
\end{figure*}

In the original TEBD$^2$ algorithm the horizontal gates are applied by rotating the whole isoTNS 90$^\circ$ four times and therefore sweeping though the isoTNS twice as often, which can lead to more errors during the MMs. Our variation on the other hand increases the numerical costs from $\mathcal{O}(d^2\chi^7+\chi^7)$ to $\mathcal{O}(d^3\chi^8+d\chi^9)$ for the MM and from $\mathcal{O}(d^2\chi^5+d^2\chi^3)$ to $\mathcal{O}(d^4\chi^8+d^6\chi^6)$ for the TEBD step. For both options it is favorable to apply the Trotter gates to the reduced tensors after judicious QR decompositions~\cite{Lubasch2014}.
We find that the double-column algorithms gives better results, while the complexity is still better than for other two-dimensional TNS algorithm with approximate contractions of the environment. For instance, the full update in PEPS requires a cost of $\mathcal{O}(d\chi^6\chi_{MPO}^2+\chi^4\chi_{MPO}^3)$ to construct a boundary MPO with bond dimension $\chi_{MPO}$, where empirically $\chi_{MPO}\propto \chi^2$ suffices. For the tensor update the construction of the reduced tensors then has an additional complexity of $\mathcal{O}(d^4\chi^4\chi_{MPO}^2+d^2\chi^6\chi_{MPO}^2+d^2\chi^4\chi_{MPO}^3)$  and $\mathcal{O}(d^6\chi^6)$ for its decomposition~\cite{Lubasch2014a}.

In Ref.~\cite{Zaletel2020} the TEBD$^2$ algorithm was successfully applied with isoTNS to find ground states with imaginary-time evolution and in Ref.~\cite{Lin2021} for spectral functions with real-time evolution. In this work we use imaginary-time evolution together with a purification ansatz to simulate thermal states.

The purification ansatz is a well known tool from MPS, where it was used for thermal state simulations~\cite{Verstraete2004, Feiguin2005, Barthel2009}. Let us consider any mixed (thermal) state described by the density matrix~$\rho$. If we enlarge the physical Hilbert space $\mathcal{H}^p$ with a copy called the ancilla Hilbert space $\mathcal{H}^a$, we can regain our physical density matrix by tracing out the ancilla legs of a pure state $\ket{\psi} \in \mathcal{H}^p \otimes \mathcal{H}^a$,
\begin{equation}\label{eq::trace}
\rho = \mathrm{Tr}_a \ket{\psi}\bra{\psi}.
\end{equation}
In terms of a TN we can think of the purification $\ket{\psi}$ as adding an additional leg on each tensor, that corresponds to the ancilla space [see Fig.~\ref{fig::1}~(b)]. When computing observables, we trace out that leg by contracting with the corresponding ancilla leg of the conjugated tensor.

To obtain thermal states, we start with a purified state at infinite temperature. For a spin-1/2 model it is given as a product of local Bell states entangling spin up and spin down of the physical and ancilla degrees of freedom,
\begin{equation}\label{eq::inf_temperature}
	\ket{\psi(\beta=0)} = \left[\frac{1}{\sqrt{2}} \left(\ket{\uparrow}_p\ket{\downarrow}_a -\ket{\downarrow}_p\ket{\uparrow}_a\right)\right]^{\otimes N}
\end{equation}
such that $\rho(\beta=0) = \mathbb{Id}$.

For any other temperature $\beta > 0$ the thermal state is then given by the imaginary-time evolution of the infinite temperature state.
\begin{equation}
	\ket{\psi(\beta)} = e^{-\frac{\beta}{2} H} \ket{\psi(\beta=0)}
\end{equation}
Here $H$ is only acting on the physical legs and we therefore obtain $\rho(\beta) \propto \mathrm{Tr}_a \ket{\psi(\beta)}\bra{\psi(\beta)} \propto e^{-\beta H}$, up to a normalization constant.
By using the TEBD$^2$ algorithm with imaginary-time evolution we can thus construct thermal isoTNS using purification. The cost of using a purification ansatz is a larger local Hilbert space dimension $d\rightarrow d^2$. For the double-column Moses Move this will increase the complexity to $\mathcal{O}(d^6\chi^8+d^2\chi^9)$, while for the TEBD a previous QR decomposition~\cite{Lubasch2014} gives a scaling $\mathcal{O}(d^6\chi^8+d^6\chi^6)$.

Note that the trace in eq.~\eqref{eq::trace} allows us to apply additional unitary operators to the ancilla legs of the pure state, without changing the thermal density matrix. This unitary degree of freedom can be utilized to decrease the entanglement of the state~\cite{Hauschild2018}. In practice, we find that obtaining the optimal disentangler on the ancilla legs is a challenging task. We compare different disentangling schemes in the Appendix~\ref{app::disentanglers}, but do not find much improvement for our results. Thus in the following we do not exploit this unitary degree of freedom on the ancilla legs.

\begin{figure*}[t]
\centering
\includegraphics[trim={0cm 0cm 0cm 0cm},clip,width=1.0\linewidth]{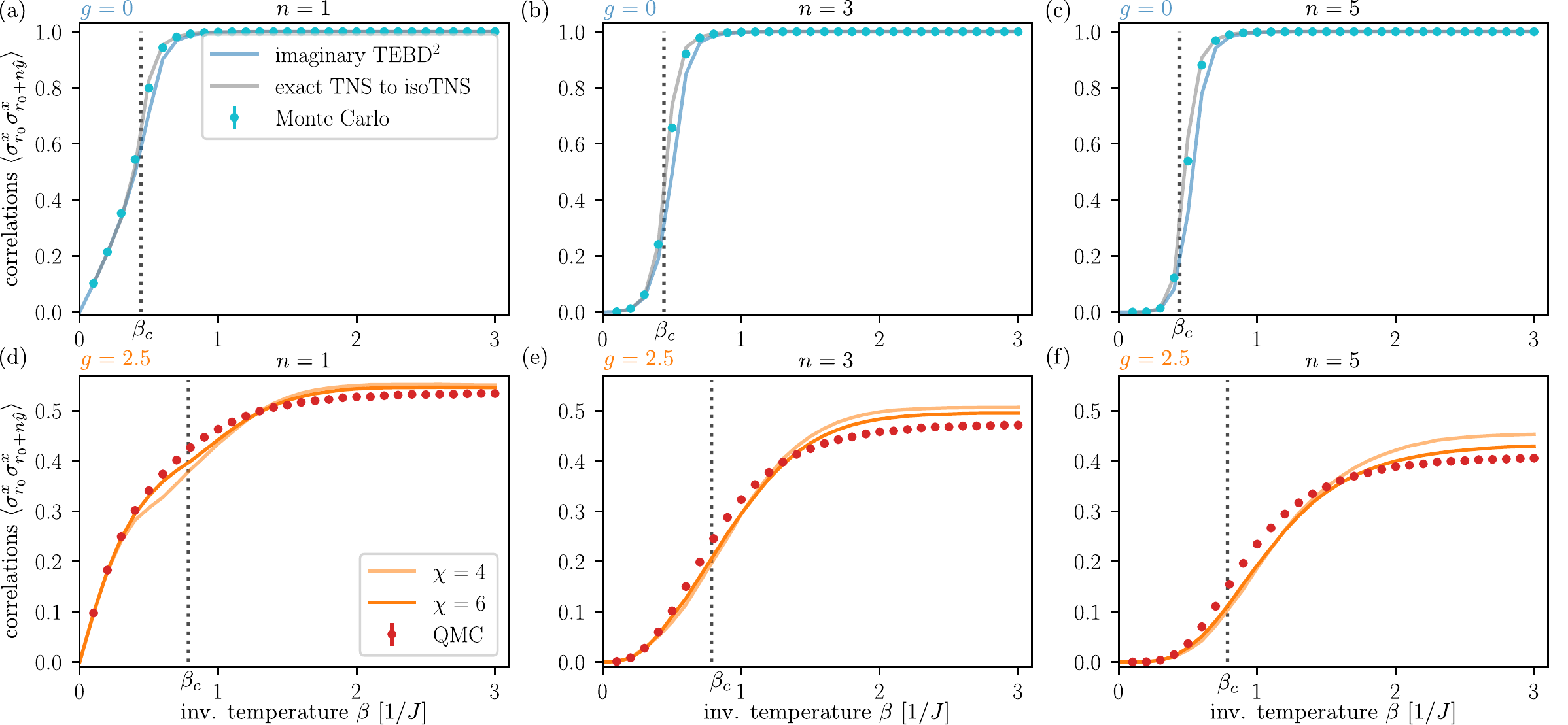}
\caption{\textbf{Correlation functions.} With isoTNS we can efficiently compute $\langle \sigma^x_{r_0} \sigma^x_{r_0+n\hat{y}} \rangle$ along the orthogonality column. Here, $r_0$ is chosen to be the fifth column and third row of a $10\times 10$ system. We compare (a),~(d) nearest neighbor, $n=1$, as well as (b,~e) $n=3$, and (c),~(f) $n=5$ correlations along the $y$-direction with (quantum) Monte Carlo results. The upper row shows data for $g=0$ from the TEBD$^2$ algorithm and from exact TNS to isoTNS conversion, each with $\chi=4$. For $g=2.5$ data for maximal bond dimensions $\chi=4$ and $6$ is compared in the bottom row. The system size is $L=10$.}
\label{fig::4}
\end{figure*}

\section{Results}\label{sec::results}
We use the previously described algorithm to obtain thermal states for the transverse field Ising model,
\begin{equation}
H = - J \left( \sum_{\langle i, j \rangle} \sigma^x_i \sigma^x_j -g \sum_{i} \sigma^z_i \right),
\end{equation}
where $\langle i, j \rangle$ denotes the nearest neighbors on a square lattice with open boundary conditions and $\sigma^\alpha_i$ are the usual Pauli matrices for spin-1/2.

The finite temperature phase diagram for this model is known from quantum Monte Carlo simulations~\cite{Hesselmann2016} [see inset Fig.~\ref{fig::3}~(c) for a sketch]. At $g = 0$ we have a classical Ising model with a thermal phase transition at $T_c = 2J/\log(1+\sqrt{2}) \approx 2.269J$ from a ferromagnetic to paramagnetic phase, while for $T=0$ there is a quantum phase transition around $g_c\approx 3.044$.
We will analyze the finite-temperature behavior along two cuts in the phase diagram corresponding to $g=0$ and $2.5$ and directly compare all results to quantum Monte Carlo (QMC) results, obtained with the \texttt{ALPS} package~\cite{ALPS}. Since our model is sign-problem free, the QMC results can be seen as exact. The small error bars are obtained from the standard deviation  of $5\times 10^6$ samples. Here we focus on a fixed system size $L=10$, results for other $L$ are shown in the Appendix~\ref{app::sizeL}.

For the classical case at $g=0$ there is also another route to obtain a purified thermal state, aside from the earlier described TEBD$^2$ algorithm. In that limit, only $\sigma^x$ operators are involved in the Hamiltonian and all terms commute. Therefore, we do not have to rely on a Trotter decomposition to construct the operator $e^{-\beta H/2}$, but can directly apply the imaginary-time evolution to $\beta$ on all bonds, starting again from the infinite-temperature state eq.~\eqref{eq::inf_temperature}. Each of the bond-gates can then be decomposed via an SVD to local tensors with virtual bond dimension two. Hence, we can construct a 2D tensor network with maximal bond dimension $\chi=2$ that exactly represents the thermal state at any temperature $\beta$ [see also inset of Fig.~\ref{fig::3}~(a)]. However, this tensor network is not in an isometric form.
To bring a general TNS into isometric form, we have to perform sweeps similar to the TEBD$^2$ algorithm, but without the unitary updates. During the sweep the isometric form is established, but will accumulate errors from the MM, which will depend on the maximal bond dimension of the resulting isoTNS; here $\chi=4$ for all cases. Therefore, starting from an exact TNS representation and iteratively ``isometrizing'' the tensor network, we can draw conclusions about the representability of thermal states with isoTNS. Note that this scheme works in the classical limit and cannot be generalized to finite transverse field $g \neq 0$, where a cluster expansion of the corresponding TN operators would be necessary~\cite{Vanhecke2021}.

\subsection{Energy density and specific heat}\label{sec::energy}

We start our analysis by evaluating the energy density $\langle E \rangle/L^2$. Since the Hamiltonian only has local terms, we can readily measure this quantity during the TEBD$^2$ sweeps. The results are shown in Fig.~\ref{fig::3}. Let us first evaluate the classical $g=0$ case. We find that the energy density agrees well with the Monte Carlo data for both the imaginary-time purification (blue lines) and the isometrization from the exact TNS (gray lines) [Fig.~\ref{fig::3}~(a)]. Only around the critical temperature small deviations are visible for the TEBD$^2$ curve. To quantify those we show the difference between isoTNS and Monte Carlo energy densities in Fig.~\ref{fig::3}~(b). The maximal deviations occur close to the critical point, where the correlation length diverges and therefore correlations cannot be captured with isoTNS of low bond dimensions anymore. This is in contrast to a generic TNS which is capable of capturing power-law correlations and hence the critical point of the classical Ising transition with a low bond dimension of $\chi=2$. Due to the isometry condition, by contrast, only exponentially decaying correlations can be captured for an orthogonality center $\Psi^{[i]}$ with finite bond dimension.
Hence, the isoTNS with fixed bond dimension $\chi=4$ cannot represent the thermal state near the phase transition exactly. This is most directly seen from the exact TNS to isoTNS conversion. For temperatures below and above the critical point, however, we find very good agreement of the energies, indicating that the thermal states in those regions can be faithfully represented. 

Likewise, when constructing the isoTNS with the TEBD$^2$ algorithm, we see a clear peak in the energy density error around $\beta_c$, which extends to a broader region at lower temperatures than the one from the exact TNS. While for the latter each data point corresponds to an independent simulation of the thermal state, during the TEBD$^2$ algorithm each state in the imaginary-time evolution depends on the previous one and therefore errors from Trotterization, truncation and the Moses Move accumulate. Nevertheless, eventually the energy approaches the ground-state energy at low temperatures, where both curves show the same energy error again. The error bars shown in Fig.~\ref{fig::3}~(b) are the ones from the Monte Carlo simulations. The good agreement between the direct conversion of the TNS and the TEBD${}^2$ algorithm shows that the latter efficiently explores the variational isoTNS manifold and properly constructs the thermal state without accumulating too serious errors.

The next question we address is how well the energy density can be computed if we also consider quantum fluctuations by turning on the transverse field $g=2.5$. In Fig.~\ref{fig::3}~(d) we can clearly see some deviations for a small maximal bond dimension $\chi=2$, indicating that this bond dimension is not sufficient to capture the quantum and thermal correlations correctly. By contrast, for $\chi=4$ and $6$ the orange curves agree very well with the QMC energy. A closer look at the energy difference [Fig.~\ref{fig::3}~(e)] reveals that similarly to the classical case, the largest discrepancy can be found around the critical temperature $T_c\approx1.274J$~\cite{Hesselmann2016}. However, we find that the peak is much broader than in the $g=0$ case and the error only saturates at much lower temperatures. This is already expected from the slower approach of the energy to the ground-state energy for increasing $\beta$. Hence, the isoTNS also approaches the well representable ground state at higher $\beta$ and the influence of the accumulated Trotterization, truncation, and MM error prevails for larger temperature ranges.

Although a direct comparison of energy densities gives a good first impression of how well the purification algorithm with TEBD$^2$ works, we are ultimately interested in observables that are easier accessible experimentally. Therefore, we now focus on the specific heat that can be directly obtained from numerical derivatives of the energy.

As already argued in~\cite{Zaletel2020}, there is a competition of Trotterization and the Moses Move, when finding the optimal time step for the imaginary-time evolution. On the one hand, a smaller $\delta \tau$ reduces the Trotter error. On the other hand, this requires more iterations of the Moses Move, where in each step an additional error is introduced. We find the optimal time step that leads to the least energy deviations from QMC at around $\delta \tau \approx 0.1/J$. Since this time step is quite large, we cannot directly numerically evaluate the derivative of the energy as a function of temperature. Instead we first interpolate our energy results with cubic splines. From that we compute the derivatives to obtain the specific heat [Figs.~\ref{fig::3}~(c) and \ref{fig::3}~(f)]. The curves agree well with the measurements obtained from (quantum) Monte Carlo simulations. For $g=0$ [Fig.~\ref{fig::3}~(c)], the peaks for both the imaginary-time evolution and the isometrization are already very close to the actual critical temperature despite the finite system size. As seen before in the energy the results from the TEBD$^2$ algorithm differ a little from the Monte Carlo data, especially close after the critical point.
For the interacting case $g=2.5$ [Fig.~\ref{fig::3}~(f)], we also find a very similar picture as for the energies. For $\chi=2$ (not shown) the energy variations are too big to obtain reliable results. By contrast, the data for $\chi=4$ and $6$ agree reasonably well with the QMC, only showing some deviations for temperatures below the critical point. Note that due to strong finite size effects the peak of the specific heat is shifted substantially compared to $\beta_c$ of the infinite system, which is also consistent with the finite-size QMC data (red dots).

\subsection{Correlation functions}\label{sec::correlations}

So far we only investigated properties that were obtained from the energy measurements. Now let us also focus on correlation functions, which are measured independently. For general TNS, the exact computation of correlation functions is an exponentially hard problem, since it requires the contraction of the whole TN. The isometric conditions for isoTNS help to simplify these computations. Especially for measuring correlations within the orthogonality column the efficient MPS scheme can be used. Here we measure correlation functions along the $y$-direction $\langle \sigma^x_{r_0} \sigma^x_{r_0+n\hat{y}} \rangle$, where $r_0$ is a site in the middle of the system, specifically for the presented $10 \times 10$ system $r_0$ is the site in the fifth column and third row. Figure~\ref{fig::4} compares the correlations for $n=1$, $3$, and $5$. For the classical case, we find again very good agreement with the Monte Carlo correlations. In the same way as for energy and specific heat, the deviations are largest around the critical temperature, but eventually approach the ground-state value again at low temperatures. 
\begin{figure}
\centering
\includegraphics[trim={0cm 0cm 0cm 0cm},clip,width=0.8\linewidth]{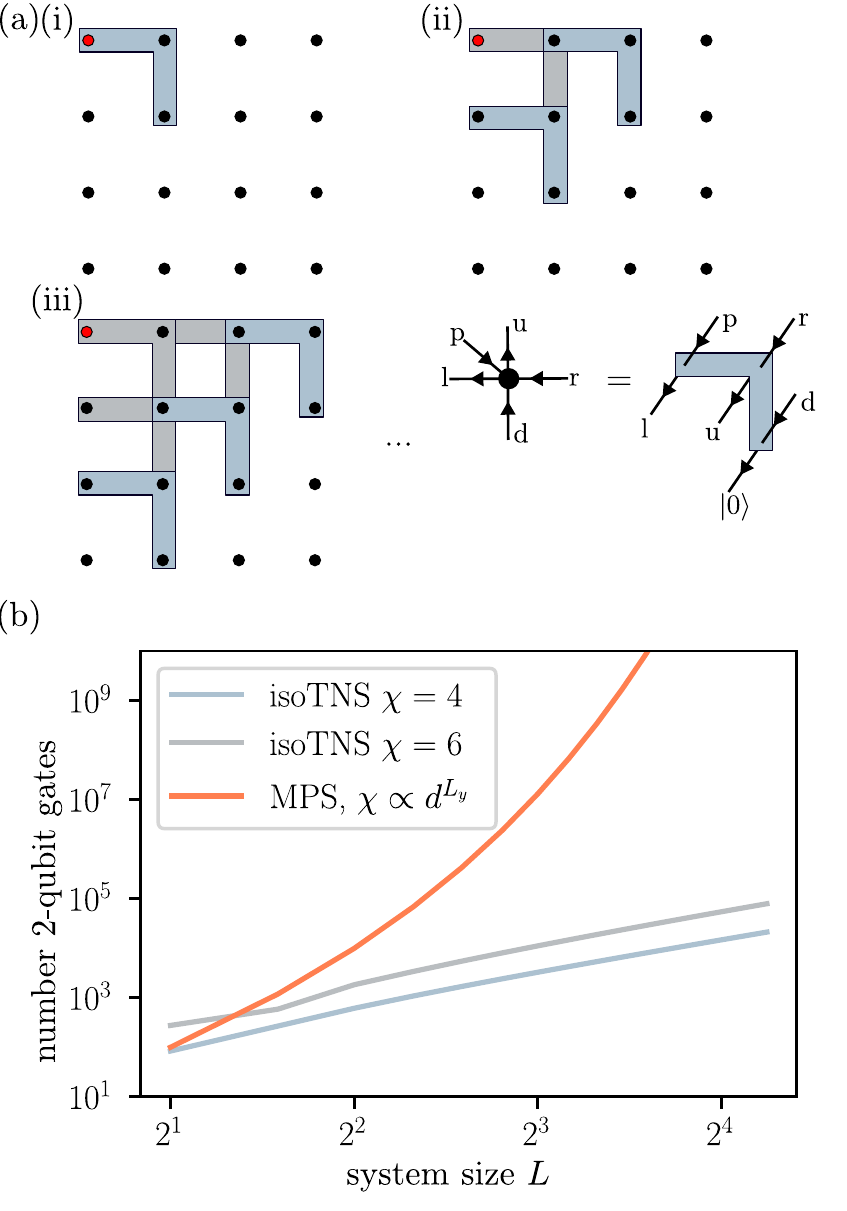}
\caption{\textbf{Quantum circuit for isoTNS.} (a) By sequentially applying gates on a two-dimensional qubit grid, one can generate an isoTNS~\cite{Wei2022}. (b) The number of required two-qubit gates for isoTNS with bond dimension $\chi$ is shown compared to representing an MPS with unitary gates, when winding the MPS around a two-dimensional lattice, which requires exponentially growing bond dimension to capture the 2D area law.}
\label{fig::5}
\end{figure}

In contrast, for $g=2.5$ the correlations even show differences to the QMC results for $T\rightarrow 0$. While these deviations are small for the nearest-neighbor correlations, they increase when considering sites that are further apart. Higher bond dimensions reduce these deviations [see Figs.~\ref{fig::4}~(c) and \ref{fig::4}~(f)]. Additionally, we find that the deviations around the critical point increase significantly, when going to larger systems indicating the absence of convergence for our accessible bond dimension (see Appendix~\ref{app::sizeL}).

These findings indicate that the correlations, especially at long distances, are in general harder to capture accurately than the local energies. While the accumulation of the errors during the TEBD$^2$ algorithm still leads to locally accurate results, the longer-range correlations become imprecise.

\section{Conclusion and Outlook}\label{sec::conclusion}

In this work, we have investigated isometric tensor network representations for thermal states of quantum many-body systems in two spatial dimensions. To this end, we apply the TEBD$^2$ algorithm to a purified isoTNS with doubled physical and ancilla space. We benchmark the algorithm with the two-dimensional transverse field Ising model, which hosts a thermal as well as a quantum phase transition. In the absence of the transverse field, we furthermore make use of the fact that the thermal state can be exactly represented as a low-bond dimension tensor network without isometry condition. Isometrizing this network provides us with insights on the variational power of the isoTNS ansatz. We found that away from the thermal phase transition at which the correlation length diverges, the isoTNS ansatz captures various properties of the state, such as the energy, specific heat, and local correlations. These results agree with the TEBD$^2$ algorithm, which is applicable for generic transverse fields. 
We also find that correlations at larger distances cannot be captured accurately with low-bond dimensions, but increasing the bond dimensions leads to more accurate results. One shortcoming of the purification algorithm is that the low-temperature states are obtained from cooling an infinite-temperature state, which leads to an accumulation of errors. This could be circumvented by variationally optimizing the R\'enyi free energy for each temperature of interest, as has been recently demonstrated for one-dimensional tensor networks~\cite{Giudice2021}.

Although in this work we focus on finite square lattices, it was also shown recently how to generalize the concept of isoTNS to infinite strips~\cite{Wu2022}. Our purification ansatz could readily be generalized to those geometries. However, the extension of isoTNS to the full two-dimensional thermodynamic limit is still an open question for future work. Other TN methods such as iPEPS or iPEPO have been explored in recent years. Although methods for these TNS have a higher complexity and the contractions of the environment can only be done approximately, they yield good results for finite temperature simulations~\cite{Czarnik2014, Czarnik2015, Czarnik2016, Czarnik2019, Kshetrimayum2019, Chen2018}.

Another important property of isoTNS is that they can be directly prepared with sequential unitary gates on quantum computers~\cite{Soejima2020,Satzinger2021, Wei2022, Slattery2021}. The sequential structure allows one to apply many local gates simultaneously, leading to a circuit depth $\mathcal{O}(L)$ [see Fig.~\ref{fig::5}~(a) for an illustration]. Furthermore, due to the structure of the isoTNS only small bond dimensions are required to capture the 2D area law. This should be contrasted to a possible MPS realization of the 2D state as a quantum circuit. In order to capture a 2D geometry, the MPS is winding around the lattice. Thus the required bond dimension to capture an area law grows exponentially with the number of sites in one spatial direction. This leads to an exponential cost when translating this MPS state to a unitary circuit. Hence, when expressing this tensor network in terms of two-qubit gates, isoTNS are suitable candidates for an efficient circuit representation as $\mathcal{O}(L^2)$ gates are required [see Fig.~\ref{fig::5}~(b)]. Therefore, isoTNS are potentially useful variational manifolds to encode thermal states on quantum computers.

\section*{Acknowledgements}
We thank J.I. Cirac, S.-H. Lin, and L. Vanderstraeten for insightful discussions. F.P. thanks in particular M. Zaletel for various collaborations and discussions related to isometric tensor networks.
We acknowledge support from the Deutsche Forschungsgemeinschaft (DFG, German Research Foundation) under Germany’s Excellence Strategy--EXC--2111--390814868 and DFG grants No. KN1254/1-2, KN1254/2-1, the European Research Council (ERC) under the European Union’s Horizon 2020 research and innovation programme (Grant Agreements No. 771537 and No. 851161), as well as the Munich Quantum Valley, which is supported by the Bavarian state government with funds from the Hightech Agenda Bayern Plus.

{\par\textit{Data availability:}} Raw data and data analysis are available on Zenodo~\cite{Zenodo}.

\appendix
\section{Disentanglers on the ancilla legs}\label{app::disentanglers}

\begin{figure}[ht]
\centering
\includegraphics[trim={0cm 0cm 0cm 0cm},clip,width=0.9\linewidth]{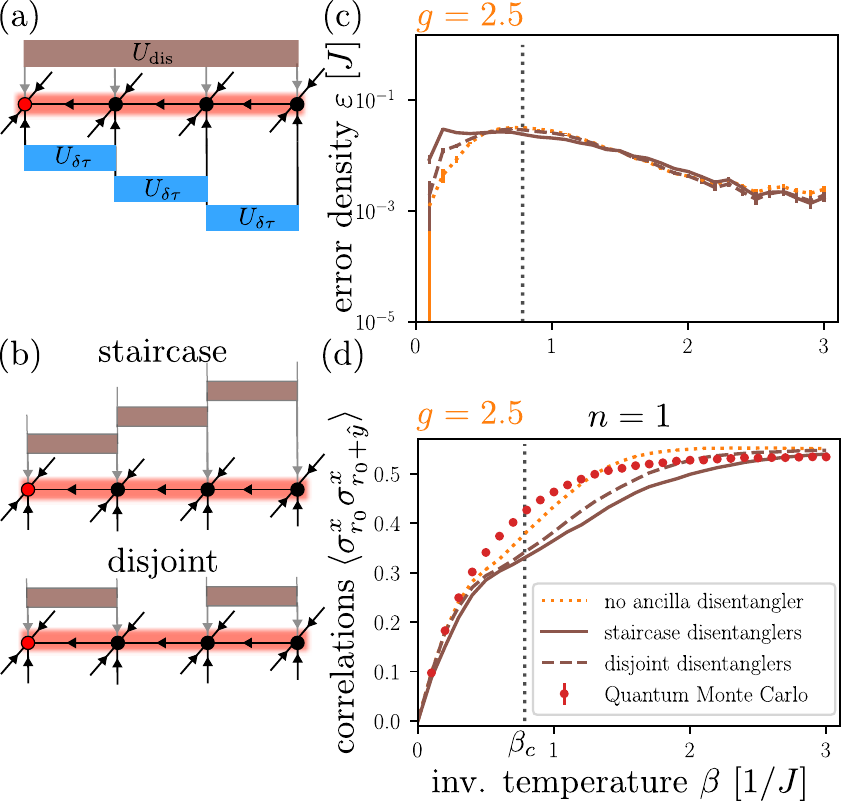}
\caption{\textbf{Disentanglers on the ancilla legs of the purified state.} (a) A unitary degree of freedom $U_{dis}$ on the ancilla legs can be utilized during imaginary-time evolution $U_{\delta \tau}$ on the physical legs. (b) We employ two-site disentanglers in a staircase pattern or a pattern where they are only applied on disjoint bonds. As in the main text we compare the error density (c) and nearest neighbor correlations (d) with and without disentanglers to quantum Monte Carlo results. We focus on system size $L=10$, maximal bond dimension $\chi=4$ and set the transverse field to $g=2.5$.}
\label{fig::S2}
\end{figure}

\begin{figure*}[ht]
\centering
\includegraphics[trim={0cm 0cm 0cm 0cm},clip,width=\linewidth]{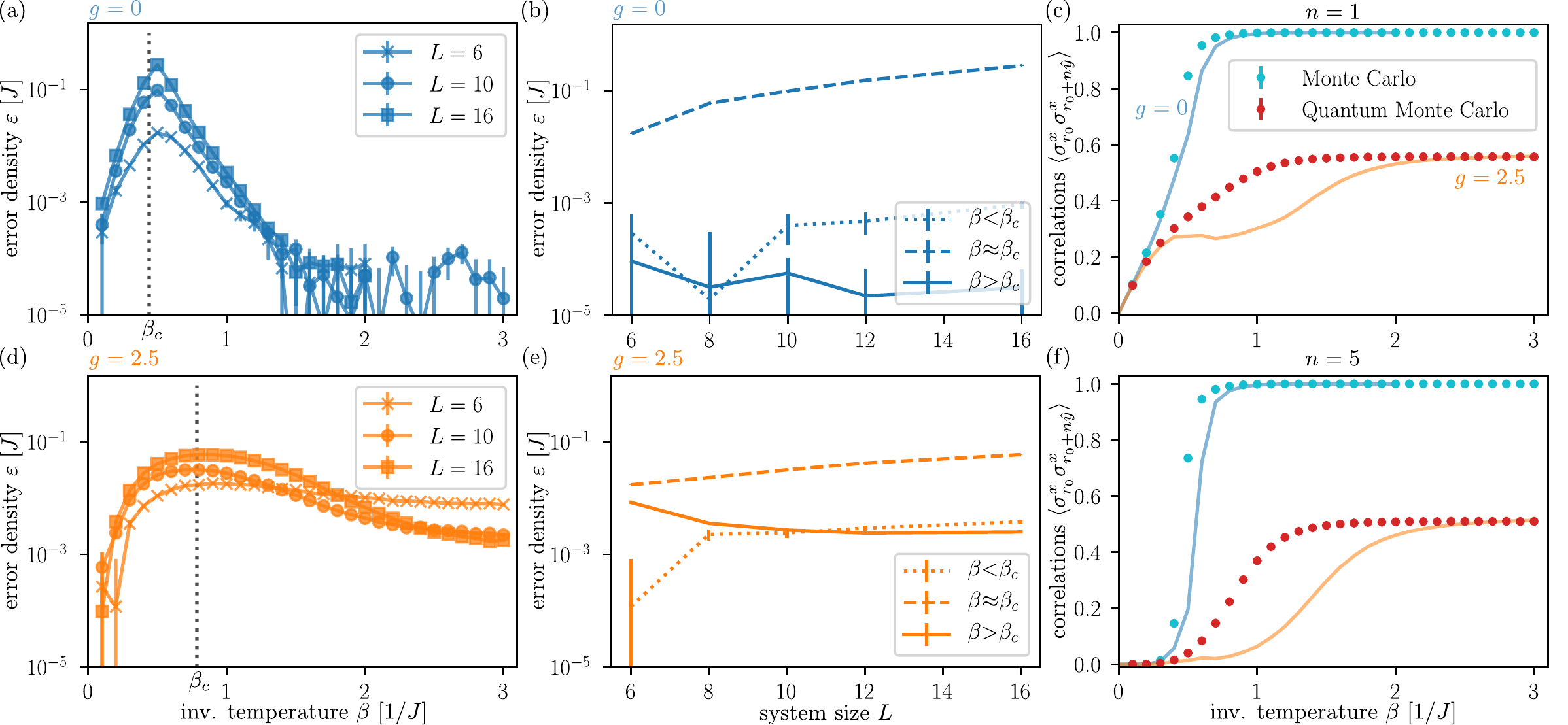}
\caption{\textbf{Error density and correlations for different system sizes.} For several system sizes between $L=6$ and $16$ we compare the error density $\varepsilon=|\langle E \rangle-E_{QMC}|/L^2$ to (quantum) Monte Carlo results for $g=0$ (blue) and $g=2.5$ (orange) as a function of inverse temperature (a,~d) as well as a function of system size $L$ for fixed $\beta=0.1/J<\beta_c$, $\beta=0.5/J\approx \beta_c$ ($\beta=0.8/J$ for $g=2.5$) and $\beta=2.5/J>\beta_c$ (b,~e). (c,~f) show $n=1$ and $n=5$ correlation functions along the $y$-direction $\langle \sigma^x_{r_0} \sigma^x_{r_0+n\hat{y}} \rangle$ for $L=16$ and $r_0$ in the 8th column and 5th row. All data shown is for fixed bond dimension $\chi=4$.}
\label{fig::S1}
\end{figure*}

In the TEBD$^2$ algorithm we have the option to utilize an additional gauge degree of freedom of the purified state. Applying unitary operators to the ancilla legs will not change the physical thermal state, but can be used to disentangle the wave function [see Fig.~\ref{fig::S2}~(a)]. In the main text, we start with a product state for infinite temperature $\beta = 0$. The imaginary-time evolution then builds up entanglement between the physical and ancilla legs throughout the whole system. However, when approaching the ground state $\beta \rightarrow \infty$, the system is again a pure state. Thus, the purified state can be written as a tensor product of the ground state (GS) with any product state (PS) on the ancilla legs $\ket{\psi}_{\beta \rightarrow \infty} = \ket{\mathrm{GS}}_p \otimes \ket{\mathrm{PS}}_a$. However, the imaginary-time evolution will in general also build up entanglement between the ancilla legs, which in turn limits the representability for a fixed bond dimension. Therefore, unitaries on the ancilla legs can be used to disentangle the ancillas of the purified state. 

For MPS this idea was systematically investigated in~\cite{Hauschild2018}, where it was shown, that a lower entanglement entropy of the state at large $\beta$ can be achieved by applying disentanglers. There two-site and global disentanglers were studied for different optimization measures.
Here we focus on two-site disentanglers that optimize the second R\'enyi entropy. The disentangling unitaries are applied on the orthogonality column during the TEBD$^2$ algorithm. We compare two different types of disentangling schemes [see Fig.~\ref{fig::S2}~(b)]. For the staircase pattern we apply the disentanglers to the ancilla legs similarly to the Trotter gates for the imaginary-time evolution. Additionally, we also try disjoint disentanglers, which only act on every second bond of the orthogonality column.

To compare the results with and without disentangler, we consider the same observables as before and choose a transverse field $g=2.5$ on a $10\times 10$ system and a maximal bond dimension of $\chi=4$. For the energy density we find very similar errors for all cases [see Fig.~\ref{fig::S2}~(c)]. Although the data with disentangler shows slightly better agreement with the QMC results around the critical point and for higher values of $\beta$, we observe larger errors at small $\beta$. The similarities of all error densities can be explained by the fact, that during the Moses Move, we already use a unitary degree of freedom between virtual legs to disentangle the state.
In contrast, for the nearest-neighbor correlations [ Fig.~\ref{fig::S2}~(d)], deviations between the data with disentanglers and without are substantial. In particular, the correlations become worse if disentanglers are used. In general, finding the optimal disentangler is a difficult problem, that becomes exponentially hard, if we want to optimize for more than two sites. Therefore, the errors in the correlations might indicate, that we are not able to find the optimal disentangler and even introduce additional correlation errors by applying the non-optimal disentangling gates. 

In summary, we find that the unitary disentanglers on the ancilla legs can in principle be useful, but do not provide any advantage in practice for our case. We attribute this on the one hand to the Moses Move which is already disentangling the degrees of freedom and on the other hand to the difficulty of finding the optimal (global) disentangler.

\section{Results for different system sizes}\label{app::sizeL}

In the main text, we focus on results for system sizes of $10\times 10$. Here, we present data also for other values of system size $L \times L$. Compared to an MPS winding on a 2D lattice, isoTNS can capture the area law with a constant bond dimension, which does not need to be scaled with $L$. Here we fix the maximal bond dimension to $\chi=4$. Moreover, we note that the number of Moses Moves required to move the orthogonality center through the system increases with $L$, which has a constant contribution to the error density.

Figures~\ref{fig::S1}~(a) and \ref{fig::S1}~(d) show the error density for $L=6, 10$ and $16$ for $g=0$ and $2.5$, respectively. We can see that close to the critical points the error increases with $L$, since there the correlation length diverges and cannot be captured with isoTNS of constant bond dimension. Nevertheless, at temperatures below and above $\beta_c$ the errors all approach the same values, indicating that indeed the error density is independent of the system size. 

The system size dependence of the error density can also be described more precisely by looking at fixed values of $\beta$ below, at, and above the critical point while tuning the system size. What can be seen in Figs.~\ref{fig::S1}~(b) and \ref{fig::S1}~(e) is a constant error density for $\beta<\beta_c$ and $\beta> \beta_c$ except for finite-size effects for $L=6$. Near the critical temperature, however, we observe a steady growth with $L$. 

When looking at correlations along the $y$-direction $\langle \sigma^x_{r_0} \sigma^x_{r_0+n\hat{y}} \rangle$, we observe a similar behavior as for the energy. Figures~\ref{fig::S1}~(c) and \ref{fig::S1}~(f) depict the correlation function for $n=1$ and $5$, respectively for a $16\times 16$ system. The site $r_0$ is chosen to be in the bulk of the system. Concretely we chose the eighth column and fifth row. At low and high temperatures the correlations agree well with (quantum) Monte Carlo results. Around the critical temperatures, the deviations are much larger than for the $L=10$ results from the main text, which is especially large for the $g=2.5$ (orange) data. We also observe that the error increases for larger distances $n$. This indicates that local observables can still be captured well with isoTNS, while for longer-range correlations larger bond dimensions may be required. Alternatively to the purification algorithm which may accumulate errors while cooling the state, the R\'enyi free energy could be optimized~\cite{Giudice2021} to see whether the large deviations for finite transverse fields are a result of the restricted variational manifold or are arising from the purification algorithm itself. 

\bibliography{main.bbl}

\end{document}